\documentclass[12pt,a4paper]{article}

\usepackage{latexsym}
\usepackage{amssymb}
\usepackage{amsmath}
\usepackage{amsfonts}
\usepackage{mathrsfs}
\usepackage{cite}

\newcommand{\be}{\begin{equation}}
\newcommand{\ee}{\end{equation}}

\def\a{\alpha}
\def\b{\beta}

\def\tr{{\rm tr}}
\def\tr{{\rm tr}\,}
\def\Tr{{\rm Tr}\,}
\def\cN{{\cal N}}

\def\bea{\begin{eqnarray}}
\def\eea{\end{eqnarray}}
\def\nn{\nonumber}

\def\cN{{\cal N}}

\def\f{\frac}

\def\tr{{\rm tr}\,}

\def\nn{\nonumber}

\def\d{\delta}

\def\sB{\stackrel{\frown}{\square}}

\def\eq{\eqref}
\def\pr{\partial}
\def\nb{\nabla}

 \numberwithin{equation}{section}

\begin{document}
\begin{titlepage}

\begin{center}
\vspace{1cm} {\Large\bf Leading low-energy effective action

\vspace{0.1cm}
in $6D$, $\cN= (1,1)$ SYM theory} \vspace{1.5cm}

 {\bf
 I.L. Buchbinder\footnote{joseph@tspu.edu.ru }$^{\,a,b,c}$,
 E.A. Ivanov\footnote{eivanov@theor.jinr.ru}$^{\,c}$,
 B.S. Merzlikin\footnote{merzlikin@tspu.edu.ru}$^{\,a,e}$,
 }
\vspace{0.4cm}

{\it $^a$ Department of Theoretical Physics, Tomsk State Pedagogical
University,\\ 634061, Tomsk,  Russia \\ \vskip 0.1cm $^b$ National
Research Tomsk State University, 634050, Tomsk, Russia \\
\vskip 0.1cm $^c$
Departamento de F\'isica, ICE,Universidade Federal de Juiz de Fora,\\
Campus Universit\'ario-Juiz de Fora, 36036-900, MG, Brazil \\
\vskip 0.15cm
 $^d$ Bogoliubov Laboratory of Theoretical Physics, JINR, 141980 Dubna, Moscow region,
 Russia \\ \vskip 0.15cm
 $^e$ Division of Experimental Physics,\\
 \it Tomsk Polytechnic University, 634050, Tomsk, Russia\\
 }
\end{center}
\vspace{0.4cm}

\begin{abstract}

We elaborate on the low-energy effective action of  $6D,\,\cN=(1,1)$
supersymmetric Yang-Mills (SYM) theory in the $\cN=(1,0)$ harmonic
superspace formulation.  The theory is described in terms of
analytic $\cN=(1,0)$ gauge superfield $V^{++}$ and analytic
$\omega$-hypermultiplet, both in the adjoint representation of gauge
group. The effective action is defined in the framework of the
background superfield method ensuring the manifest gauge invariance
along with manifest $\cN=(1,0)$ supersymmetry. We calculate leading
contribution to the one-loop effective action using the on-shell
background superfields corresponding to the option when gauge group
$SU(N)$ is broken to $SU(N-1)\times U(1)\subset SU(N)$. In the
bosonic sector the effective action  involves the structure $\sim
\frac{F^4}{X^2}$, where $F^4$ is a monomial of the fourth degree in
an abelian field strength $F_{MN}$ and $X$ stands for the scalar
fields from the $\omega$-hypermultiplet. It is manifestly
demonstrated that the expectation values  of the hypermultiplet
scalar fields play the role of a natural infrared cutoff.
\end{abstract}

\end{titlepage}

\setcounter{footnote}{0}
\setcounter{page}{1}


\section{Introduction}

The low-energy effective action plays an important role in
supersymmetric gauge theories, providing a link between
superstring/brane theory and quantum field theory. On the one hand,
such an effective action can be calculated in the quantum field
theory setting and, on the other, it can be derived within the brane
stuff. As a result, the low-energy effective action allows one, in
principle, to describe the low-energy string effects by methods of
quantum field theory and vice versa (see reviews
\cite{BUCH1,BUCH2,BUCH3}).

It is known that D3-branes are related to $4D,\,\cN=4$ SYM theory
(see, e.g., \cite{GK,BKLS}). Interaction of D3-branes is described
in abelian bosonic sector by the Born-Infeld action, with the
leading low-energy correction of the form $\sim \frac{F^4}{X^4}$,
where $F^4$ is a structure of fourth degree in an abelian field
strength $F_{mn}$ and $X$ stands for the scalar fields of $4D,\,
\cN=4$ gauge (vector) multiplet. The one-loop calculation of such an
effective action in the Coulomb branch of $\cN=4$ SYM theory, both
in the component approach and in terms of  $\cN=1,2$ superfields,
has been performed in ref.
\cite{GKPR,PvU,G-rR,BBKO,BK,BBK,LvU,BKT,KM,Kuz01}. The complete
$\cN=4$ structure of the one-loop low-energy effective action has
been established in \cite{BI,BIP}. The two-loop contributions to the
low-energy effective actions of $\cN=4$ SYM theory have been studied
in \cite{BPT,KU}. The structure of the low-energy effective action
in the mixed Coulomb - Higgs branch was a subject of ref.
\cite{BP07}. A review of the results related to the calculations of
low-energy effective actions in four-dimensional extended
supersymmetric gauge theories can be found, e.g.,  in
\cite{BUCH1,BUCH2}.

Another interesting class of the extended objects in
superstring/brane theory is presented by D5-branes (see e.g.,
\cite{GK,BKLS}). These objects are related to $6D,\cN=(1,1)$ SYM
theory likewise  D3-branes are related to $4D,\,\cN=4$ SYM theory.
Similarly to the D3-brane case, the interaction of D5-branes is
described by the $6D$ Born-Infeld action \cite{T} (see
\cite{Sevrin1,Sevrin2,Sevrin3,Sevrin4,DHHK,GTKK} for aspects of the
Born-Infeld action in diverse dimensions). Since D5-brane is related
to $6D,\cN=(1,1)$ SYM theory, it is natural to expect that the
D5-brane interaction in the low-energy domain can be calculated on
the basis of the low-energy quantum effective action of this theory.

In the present paper we study the quantum aspects of
$6D,\,\cN=(1,1)$ SYM theory. It is the maximally extended
supersymmetric gauge model in six dimensions, and it involves eight
left-handed and eight right-handed supercharges. An equal number of
spinors with mutually-opposite chiralities guarantees  the absence
of chiral anomaly in the theory. From the point of view of $6D,\,
\cN=(1,0)$ supersymmetry, the model is built on a gauge (vector)
multiplet and a hypermultiplet. Respectively, the  bosonic sector of
the model includes a real vector gauge field and two complex (or
four real) scalar fields.

Although $6D,\, \cN=(1,1)$ non-abelian SYM theory is
non-renormalizable by power counting, it was proved that it is
on-shell finite at one and two loops
\cite{FT,MarSag1,MarSag2,HS,HS1,BHS,BHS1,Bork}. Moreover, it was
recently shown that this theory is one-loop finite even off-shell
\cite{BIMS-a,BIMS-b,BIMS-c} and that the two-loop diagrams with
hypermultiplet legs are also off-shell finite \cite{BIMS-d}.

Here we develop a method to determine the one-loop effective action
in general $6D,\cN=(1,1)$ SYM theory and to calculate the leading
low-energy contributions to it. To preserve as many manifest
supersymmetries as possible we use the harmonic superspace approach
\cite{GIKOS,GIOS}. The theory under consideration is formulated in
terms of $\cN=(1,0)$ harmonic superfields describing the gauge
multiplet and the hypermultiplet. Therefore it possesses the
manifest $\cN=(1,0)$ supersymmetry and, in addition, a non-manifest
(hidden) on-shell $\cN=(0,1)$ supersymmetry mixing $\cN=(1,0)$ gauge
multiplet and hypermultiplet. These supersymmetries close on the
total on-shell $\cN=(1,1)$ supersymmetry. Such a formulation of
$\cN=(1,1)$ SYM theory was described in detail in the paper
\cite{BIS} (see also ref. \cite{HSW,Z}) \footnote{There is also
another superfield formulation for maximally supersymmetric
Yang-Mills theories based on the pure spinor superfield formalism
\cite{C1,C2}. However, the scheme of quantum calculations within
this approach has not been worked out so far.}. An essential
difference of our consideration here is the use of the so called
``$\omega$-form'' of the hypermultiplet (see below).

The theory under consideration is quantized in the framework of
$\cN=(1,0)$ supersymmetric background field method
\cite{BIMS-b,BIMS-c}. In this method, the effective action depends
on the background superfields of $6D, \,\cN=(1,0)$ gauge multiplet
and hypermultiplet. By construction, it exhibits manifest gauge
invariance under the classical gauge transformations and $\cN=(1,0)$
supersymmetry. To calculate the one-loop effective action we make
use of the superfield proper-time technique \cite{book}, which
ensures the manifest gauge invariance and $\cN=(1,0)$ supersymmetry
at all steps of calculation. The low-energy effective action is
obtained, when we impose the restriction that both the background
superfield strength and the background hypermultiplet are
space-time-independent. The leading low-energy approximation amounts
to keeping those terms in the effective action which are of the
lowest order in the superfield strength. We also assume that the
background superfields satisfy the classical equations of motion,
that guarantees the gauge independence of the effective action.

We consider the case when gauge symmetry $SU(N)$
is broken to $SU(N-1)\times U(1) \subset SU(N)$. Technically, this
means that background superfields align through the fixed generator
of Cartan subalgebra of $SU(N)$,  which corresponds to an abelian
subgroup $U(1)$. In this case the effective action of the theory
depends only on the abelian vector multiplet and hypermultiplet. In the
bosonic sector we find out the effective action for the single real $U(1)$
gauge field and four real scalar fields. The same number of bosonic
world-volume degrees of freedom is needed to describe  a single
D5-brane in six dimensions \cite{S_98}.

The paper is organized as follows. In section 2 we formulate an
arbitrary $6D,\, \cN=(1,1)$ SYM theory in terms of $\cN=(1,0)$
harmonic superfields representing the gauge multiplet and the
hypermultiplet. Unlike the majority of the previous papers on
effective action in $4D$ and $6D$ harmonic superspaces, we prefer to
work with $\omega$-form of the hypermultiplet. Such a formulation
has certain merits over the more accustomed formulation in terms of
$q$-hypermultiplet. Although the $\omega$- and $q$- descriptions of
the hypermultiplet are classically equivalent \cite{GIOS}, and this
equivalency apparently extends to the exact quantum theory, the
approximate schemes for calculating the quantum effective action in
terms of these superfields can be different. Besides, the
$\omega$-hypermultiplet possesses an advantage of being real, {\it
i.e.} carrying no external  $U(1)$ charges. This property
essentially simplifies the construction of the super-invariants in
the $\omega$-representation. In section 3, besides giving details of
the $6D, \cN=(1,1)$ SYM action in terms of the harmonic superfields
$V^{++}$ and $\omega$, we derive the transformations of the hidden
on-shell ${\cal N}=(0,1)$ supersymmetry which mixes the superfields
$V^{++}$ and $\omega$ and leaves the action invariant.  In section 3
we develop a procedure of constructing the one-loop effective action
which depends on both the gauge and the hypermultiplet background
superfields. Also we demonstrate advantages of the
$\omega$-hypermultiplet formulation and construct some on-shell
invariants depending on $V^{++}$ and $\omega$ superfields. Analogous
invariants have never been constructed in terms of
$q$-hypermultiplet. Section 4 describes the calculation of leading
low-energy contributions to the one-loop effective action. To this
end, we fix the background superfields by requiring them to be
space-time independent and to satisfy the classical equations of
motion.  We consider the case of background superfields breaking
$SU(N)$ gauge group of the original Lagrangian to $SU(N-1) \times
U(1)$. In this case the effective action depends on an abelian
$U(1)$ gauge superfield. It is explicitly demonstrated that the
$\omega$-hypermultiplet acts as an infrared regulator securing the
absence of the infrared singularities in the low-energy effective
action. The last section contains a brief summary of the results
obtained and a list of some problems for the future study.


\section{The model and conventions}

We consider the formulation of $6D, \,\cN=(1,1)$ SYM theory in terms
of $6D,\, \cN=(1,0)$ harmonic analytic superfields $V^{++}$ and
$\omega$, which represent  the gauge multiplet and the
hypermultiplet\footnote{ These superfields satisfy the Grassmann
harmonic analyticity conditions $D^{+}_{a}V^{++}=0$ and $
D^{+}_{a}\omega=0$, where $D^+_a = \tfrac{\partial}{\partial
\theta^{-a}} $.}. The action of $\cN = (1,1)$ SYM theory is written
as
 \bea
S_0[V^{++}, q^+]&=&
\frac{1}{\rm f^2}\Big\{\sum\limits^{\infty}_{n=2} \frac{(-i)^{n}}{n} \tr \int
d^{14}z\, du_1\ldots du_n \frac{V^{++}(z,u_1 ) \ldots
V^{++}(z,u_n ) }{(u^+_1 u^+_2)\ldots (u^+_n u^+_1 )}  \nn \\
&& - \f12 \tr \int d\zeta^{(-4)}\, \nb^{++}\omega \nb^{++}\omega
\Big\}\,,\label{S0}
 \eea
where ${\rm f}$ is a dimensionful coupling constant ($[{\rm f}]=-1$)
and the measure of integration over the analytic subspace
$d\zeta^{(-4)}$ includes the integration over harmonics,
$d\zeta^{(-4)} = d^6 x_{(\rm an)}\, du\,(D^-)^4$. Both $V^{++}$ and
$\omega$ superfields take values in the adjoint representation of
the gauge group. The covariant harmonic derivative $\nb^{++}$ acts
on the hypermultiplet $\omega$ as
 \be
\nabla^{++}\omega = D^{++} \omega + i [V^{++},\omega]\,. \label{Vfirst}
 \ee
The action \eq{S0} is invariant under the infinitesimal gauge transformations
 \bea
 \d V^{++} = -\nb^{++} \Lambda\,,  \quad \d \omega =  i[\Lambda, \omega]\,,
 \label{gtr}
 \eea
where $\Lambda(\zeta, u) = \widetilde{\Lambda}(\zeta, u)$ is a real
analytic gauge parameter.

Besides the analytic gauge connection $V^{++}$ we introduce a non-analytic
one $V^{--}$ as a solution of the zero curvature condition \cite{GIOS}
 \bea
 D^{++} V^{--} - D^{--}V^{++} + i [V^{++},V^{--}]=0\,. \label{zeroc}
 \eea
Using $V^{--}$ we can define one more covariant harmonic
derivative $\nb^{--} = D^{--} + i V^{--}$ and the ${\cal
N}=(1,0)$  gauge superfield strength
 \bea
 W^{+a}=-\f{i}{6}\varepsilon^{abcd}D^+_b D^+_c D^+_d V^{--}\,,
 \eea
possessing the useful off-shell properties
 \bea
\nabla^{++} W^{+a}  = \nabla^{--} W^{-a} \ =\ 0\,, \qquad W^{-a} =
\nabla^{--}W^{+a} \,. \label{HarmW}
 \eea

Introducing an analytic superfield $F^{++}\,$,
 \bea
 F^{++} = \f14 D^{+}_a W^{+ a} = i(D^+)^4 V^{--}\,, \qquad D^+_a F^{++} = \nb^{++} F^{++}=0\,,
 \eea
we can write the classical equations of motion corresponding to the action \eq{S0} as
 \bea
 F^{++}  +[\omega,\nabla^{++}\omega] = 0\,, \qquad  (\nabla^{++})^2\,\omega = 0\,.
\label{Eqm}
 \eea

The $\cN=(1,0)$ superfield action \eq{S0} enjoys the additional $\cN=(0,1)$ supersymmetry
 \bea
\delta V^{++} &=& (\epsilon^{+ A}u^+_A) \omega - (\epsilon^{+
A}u^-_A) \nabla^{++}\omega
= 2 (\epsilon^{+ A}u^+_A) \omega - \nabla^{++} \big((\epsilon^{+ A}u^-_A)
 \omega\big),\label{V++HidOm} \\
\delta \omega &=& -(D^+)^4 \big((\epsilon^{-A}u^-_A) V^{--}\big) =
i(\epsilon^{- A}u^-_A) F^{++} -i (\epsilon^A_a u^-_A)\, W^{+ a}
\label{omegaHid},
 \eea
where $A=1,2$ is the  Pauli-G\"{u}rsey $SU(2)$ index. To check this,
one first derives,  using
\eq{V++HidOm} and \eq{omegaHid},  the ${\cal N}=(0,1)$ transformation law of
$\nb^{++}\omega$
 \bea
\delta(\nabla^{++}\omega) = i\big( (\epsilon^{-A}u^+_A)  +
(\epsilon^{+A}u^-_A)\big)F^{++} -i (\epsilon^A_a u^+_A)\, W^{+ a}
+ i(\epsilon^{+ A}u^-_A)[\omega, \nabla^{++}\omega]. \label{omega3Hid}
 \eea
Then  one varies the classical action \eq{S0} with respect to \eq{V++HidOm} and
\eq{omega3Hid}
 \bea
 \d S = \f{1}{\rm f^2}  \Big\{\tr\int d^{14} z du \,V^{--} \d V^{++} -
 \tr\int d \zeta^{(-4)} \nb^{++} \omega\, \d( \nb^{++}
 \omega)\Big\}\,.
 \eea
In the first integral, we pass to the integration over the analytic subspace and use
the explicit form of the variations  \eq{V++HidOm} and \eq{omega3Hid}
 \bea
\d S &=& -\f{i}{\rm f^2}  \tr\int d \zeta^{(-4)} \Big\{ 2F^{++}
(\epsilon^{+ A}u^+_A) \omega +\nb^{++}\omega \big(
(\epsilon^{-A}u^+_A)+(\epsilon^{+A}u^-_A)\big)F^{++} \nn \\
 && -F^{++}\nabla^{++}
\big((\epsilon^{+ A}u^-_A) \omega \big) - \epsilon^A_a u^+_A\,
\nb^{++}\omega\, W^{+ a} \Big\} =0\,. \label{var}
 \eea
The last two terms in \eq{var} are the total harmonic derivative
$\nb^{++}$ due to the properties of $F^{++}$ and $W^{+a}$ and so they vanish
under $ d \zeta^{(-4)}$. The first two terms cancel each other
after integration by parts with respect to the harmonic derivative $\nb^{++}$  and
using the properties $\nb^{++} \epsilon^{-A} = \epsilon^{-A}$ and $\nb^{++} u^{-}_A
= u^{+}_A$. Finally, the term  $\tr \big( \nb^{++}\omega
[\omega, \nb^{++}\omega]\big)$ vanishes due to the cyclic property of trace.

The zero curvature condition \eq{zeroc} allows one to express the
transformation of the non-analytic gauge connection $\delta V^{--}$
through $\d V^{++}$
 \be
 \nabla^{++}\delta V^{--} - \nabla^{--}\delta V^{++} = 0\,,
 \ee
and to define the transformation low of the gauge superfield strength
$W^{+a}$ under the hidden supersymmetry
 \bea
 \delta W^{+ a} =\varepsilon^{adbc} \epsilon^A_d \nabla_{bc}
  \big(u^+_A \omega - u^-_A \nabla^{++}\omega\big)
 + i\epsilon^{- A}[W^{+ a}, u^+_A \omega - u^-_A \nabla^{++}\omega],
 \label{WHid}
 \eea
where
 \bea
\nabla_{bc} = \partial_{bc} -\frac12 D^+_b D^+_c
\,V^{--}\,.
 \eea
Note that, while deriving \eq{WHid}, we essentially used the
$\omega$-hypermultiplet equation of motion $(\nabla^{++})^2 \omega =
0\,$ and some its consequences.


\section{One-loop effective action in the  background field method}

The background field method for $4D, {\cal N}=2$ gauge theories in
the harmonic superspace was worked out in \cite{BBKO}. It was
generalized to $6D$ theories in our recent works
\cite{BIMS-b,BIMS-c}). Following these techniques,   we represent
the original superfields $V^{++}$ and $\omega$ as a sum of the
``background'' superfields ${\bf V}^{++}, {\bf \Omega}$ and the
``quantum'' ones $v^{++}, \omega\,$,
 \be
 V^{++}\to {\bf V}^{++} + {\rm f} v^{++}, \qquad \omega \to {\bf \Omega} + {\rm f} \omega\,,
 \ee
and then expand the action in a power series with respect to the
quantum fields. The one-loop contribution to the effective action
$\Gamma^{(1)}$ for the model \eq{S0} is given by
 \be
 e^{ i\Gamma^{(1)}[{\bf V}^{++}, {\bf \Omega}]} =\mbox{Det}^{1/2}\sB \int {\cal
D}v^{++}\,{\cal D}\omega\, {\cal D}{\bf b}\,{\cal D}{\bf c}\,{\cal
D}\varphi\,\,\, e^{iS_{2}[v^{++}, \omega, {\bf b}, {\bf c}, \varphi,
{\bf V}^{++}, {\bf \Omega}]}\,,
 \label{Gamma0}
 \ee
where
  \bea
 S_2 &=& S_{gh}+ \frac{1}{2}\tr\int d\zeta^{(-4)}\, v^{++}\sB v^{++}
  -\f12 \tr\int d\zeta^{(-4)}\,  (\nb^{++} \omega)^2 \nn \\
 && - i\tr \int d\zeta^{(-4)}\Big\{
  \nb^{++}\omega [v^{++},{\bf \Omega}]
  + \nb^{++}{\bf \Omega} [v^{++},\omega] +\f{i}2[v^{++},{\bf \Omega}]^2\Big\} \,, \label{S2}\\
  S_{gh} &=& \tr \int d\zeta^{(-4)}\,{\bf b}(\nb^{++})^{2}{\bf c}
 + \frac{1}{2}\tr\int
 d\zeta^{(-4)}\,\varphi(\nb^{++})^{2}\varphi\,. \label{Sgh}
 \eea
The ghost action $S_{gh}$ \eq{Sgh} involves the Faddeev-Popov ghosts
${\bf b}$ and ${\bf c}$ and  also Nielsen-Kallosh ghost $\phi$. The
covariantly-analytic d'Alembertian $\sB$ is defined as
$\sB=\frac{1}{2}(D^+)^4(\nb^{--})^2$, where the harmonic covariant
derivative $\nb^{--} = D^{--} + i {\bf V}^{--}$ contains the
background superfield ${\bf V^{--}}$. While acting on an analytic superfield, the operator
$\sB$ is given by
\begin{eqnarray}
&& \sB= \eta^{MN} \nabla_M \nabla_N + {\bf W}^{+a} \nabla^{-}_a +
{\bf F}^{++} \nabla^{--} - \frac{1}{2}(\nabla^{--} {\bf F}^{++})\,,
\label{Box_First_Part}
\end{eqnarray}
where $\eta^{MN} = {\rm diag} (1, -1, -1, -1, -1, -1)$ is the six-dimensional Minkowski metric,
$M,N=0,..,5$, and $\nabla_{M}=\pr_M + i{\bf A}_M$ is the background-
dependent vector supercovariant derivative (see \cite{BIS} for details).

In the action \eq{Gamma0} the background superfields ${\bf V^{++}} $
and ${\bf \Omega}$ are analytic but unconstrained otherwise.
The gauge group of the theory \eq{S0} is assumed to be $SU(N)$. For the further
consideration, we will also assume that the background fields ${\bf
V}^{++}$ and  ${\bf \Omega}$ align in a fixed direction in the
Cartan subalgebra of $su(N)$
 \bea
 {\bf V}^{++} = V^{++}(\zeta,u) H\,, \qquad {\bf \Omega} = \Omega(\zeta,u)\, H\,,
 \eea
where $H$ ia a fixed generator in  the Cartan subalgebra generating
some abelian subgroup $U(1)$ \footnote{We denote the $H$ component
of ${\bf V}^{++}$ by the same letter $V^{++}$ as the original
non-abelian harmonic connection, with the hope that this will not
create a misunderstanding. The same concerns the abelian superfield
strength $W^{+ a}$.}. Our choice of the background corresponds to
the spontaneous symmetry breaking $SU(N) \rightarrow SU(N-1)\times
U(1)$. We have to note that the pair of the background superfields
$(V^{++},\Omega)$ forms an abelian vector $\cN=(1,1)$ multiplet
which, in the bosonic sector, contains  a single real gauge vector
field $A_M(x)$ and four real scalars $\phi(x)$ and
$\phi^{(ij)}(x)\,, i,j=1,2$, where $\phi$ and $\phi^{(ij)}$ are the
scalar components of $\Omega$ hypermultiplet \cite{GIOS}. The
abelian vector field and four scalars in six-dimensional space-time
describe just the bosonic world-volume degrees of freedom of a
single D5-brane \cite{GK,BKLS}.

The classical equations of motions \eq{Eqm} for the background
superfields $V^{++}$ and  $\Omega$ are free
 \bea
 F^{++} = 0\,, \qquad (D^{++})^2 \Omega = 0\,. \label{EqmB}
 \eea
In that follows we assume that the background superfields solve the
classical equation of motion \eq{EqmB}. We will also consider the
background slowly varying in space-time, {\it i.e.}  assume that
 \bea
 \partial_M W^{+a} = 0\,, \qquad \partial_M \Omega = 0\,.
 \label{constBG}
 \eea

Finally we are left with an abelian background analytic superfields  $V^{++}$
and $\Omega$, which satisfy the classical equation of motion
\eq{EqmB} and the conditions \eq{constBG}. Under these assertions the
gauge superfield strength $W^{+a}$ is analytic \footnote{ In
general this is not true and $F^{++}\neq0$.}, $D^+_a W^{+b} =
\d_a^b F^{++}=0$. For further analysis it is convenient to use the
$\cN=(0,1)$ transformation for gauge superfield strength $W^{+a}$
\eq{WHid}. In the case of the slowly varying abelian on-shell
background superfields the hidden $\cN=(0,1)$ supersymmetry
transformations \eq{omegaHid} and \eq{WHid} have the very simple form
 \bea
  \delta \Omega = -i (\epsilon^A_a u^-_A)\, W^{+ a}\, \qquad \delta W^{+ a}=0.
  \label{Onshell2}
 \eea
These transformation rules follow from the abelian version of the
transformations  \eq{omegaHid}, \eq{WHid} in which one should take
into account the conditions  \eq{constBG} and \eq{EqmB}. It is worth
to point out that these conditions  on their own are covariant under
${\cal N}=(0,1)$ supersymmetry.

In conclusion of this section, let us consider the simplest $\cN=(1,1)$
invariants which can be constructed out of  the abelian analytic superfields
$W^{+a}$ and $\Omega$ under the assumptions \eq{EqmB} and
\eq{constBG}. It is evident that the following gauge-invariant
action
 \bea
I={\rm f}^2 \int d\zeta^{(-4)} (W^{+})^4 {\cal F}({\rm f}\Omega),
\label{I}
 \eea
where $(W^+)^4 =
-\f{1}{24}\varepsilon_{abcd} W^{+a}W^{+b}W^{+c}W^{+d}$ and ${\cal
F}({\rm f}\Omega)$ is an arbitrary function of $\Omega$,
is invariant under the transformation \eq{Onshell2} due to the
nilpotency condition $(W^{+})^5\equiv0$. For our further consideration, of the main interest
is the choice
 \bea
I_1= c\int d\zeta^{(-4)}\, \f{(W^{+})^4}{ \Omega^2}\,, \label{I1}
 \eea
which corresponds to ${\cal F} = \f{1}{{\rm f^2}\Omega^2}$ in
\eq{I}. The coefficient $c$ in \eq{I1} cannot be fixed only on the
symmetry grounds and should be calculated in the framework of the
quantum field theory. In the next section we will find it from the calculation of the
leading low-energy contribution to the effective action of the theory \eq{S0}.


\section{Leading low-energy contributions to one-loop effective action}

We  choose the Cartan-Weyl basis for the $SU(N)$ gauge group generators,
so that the quantum superfield $v^{++}$ has the decomposition
 \bea
 v^{++} =  v^{++}_{\rm i} H_{\rm i}+ v^{++}_\a E_\a\,, \qquad {\rm i} = 1,..,
 N-1,\quad \a = 1,..,N(N-1)\,,
 \eea
where $E_\a$ is the generator corresponding to the root $\a$
normalized as $\tr(E_\a E_{-\b} )=  \d_{\a\b}$ and $H_{\rm i}$ are
the Cartan subalgebra generators, $[H_{\rm i}, E_\a] = \a_{H_i}
E_\a$. In this case the background covariant d'Alembertian
\eqref{Box_First_Part} under the conditions \eq{EqmB} acts on the
quantum superfield $v^{++}$ as
 \bea
 \sB v^{++}&=& \f12 (D^+)^4 \Big\{ (D^{--})^2 v^{++}
 + i \a_H D^{--} V^{--}v_\a^{++} E_\a
 \nn \\ &&  \qquad\qquad + i \a_H V^{--} D^{--}
 v^{++}_\a E_\a  - \a^2(H) (V^{--})^2 v^{++}_\a E_\a \Big\} \\
  &=& \sB_H\, v^{++}_\a E_\a + \pr_M\pr^M\, v^{++}_{\rm i} H_{\rm i}
  \,,
 \eea
where we have introduced the operator
 \bea
 \sB_{H} := \square + \a_{H}\, W^{+a} D^-_a \,. \label{sboxH}
 \eea

The one-loop effective action \eq{Gamma0} for the background superfields
$V^{++}$ and $\Omega$ subjected to the conditions \eq{EqmB} and \eq{constBG} thus reads
 \bea
\Gamma^{(1)}&=& \f{i}2\Tr_{(2,2)} \ln\Big(\sB_H - \alpha_H^2
\Omega^2\Big) + \f{i}2 \Tr \ln\Big[(\nb_H^{++})^2 +
A_{(+)}\f{\alpha^2_H}{\sB_H - \alpha_H^2 \Omega^2} A_{(-)}\Big]   \nn \\
&& - \f{i}2\Tr_{(4,0)} \ln \sB_H -i\Tr\ln (\nb_H^{++})^2 + \f{i}2
\Tr\ln (\nb_H^{++})^2\,, \label{1loop}
 \eea
where we have defined $\nb_H^{++} = D^{++} + \alpha_H V^{++}$ and $
 A_{(\pm)}(\Omega) = \Omega \nb_H^{++} \pm \tfrac32 (D^{++}\Omega)$.

The first term in the first line of the expression \eq{1loop} is the
contribution from the gauge multiplet, while the second one is the
total contribution from the hypermultiplet.  The first term in the
second line comes from $\mbox{Det}^{1/2}\sB $ in \eq{1loop}, while
the second and the third ones are contributions from the ghost
action \eq{Sgh}.  We use the standard definition for the functional
trace over harmonic superspace in \eq{1loop}
$$\Tr_{(q,4-q)} {\cal O} = \tr \int d \zeta_1^{(-4)}d \zeta_2^{(-4)}
\, \d_{\cal A}^{(q,4-q)}(1|2)\,  {\cal O}^{(q,4-q)}(1|2)\,.
$$
Here $ \d_{\cal A}^{(q,4-q)}(1|2)$ is an  analytic delta-function
\cite{GIOS} and  ${\cal O}^{(q,4-q)}(\zeta_1,u_1|\zeta_2,u_2)$ is
the  kernel of an operator acting in the space of analytic
superfields with the harmonic U(1) charge $q$.

As the next step we rewrite the contribution from $\mbox{Det}^{1/2}\sB $
as
 \bea
 \f{i}2\Tr_{(4,0)}\ln\sB_H = \f{i}2\Tr_{(4,0)}\ln \Big(\sB_H - \alpha^2_H
 \Omega^2\Big) + \f{i}2\Tr_{(4,0)}\Big(1+\f{\alpha^2_H\Omega^2}{\sB_H - \alpha^2_H
 \Omega^2}\Big)\,. \label{det}
 \eea
Hence the one-loop contribution to effective action \eq{1loop} is divided as
 \bea
 \Gamma^{(1)} = \Gamma^{(1)}_{\rm lead} + \Gamma^{(1)}_{\rm high}\,,
 \eea
where
 \bea
 \Gamma^{(1)}_{\rm lead} = \f{i}2\Tr_{(2,2)} \ln\Big(\sB_H - \alpha_H^2
 \Omega^2\Big) - \f{i}2\Tr_{(4,0)} \ln\Big(\sB_H -
 \alpha_H^2\Omega^2\Big), \label{1loop2}
 \eea
and
 \bea
\Gamma^{(1)}_{\rm high}&=& \f{i}2 \Tr \ln\Big[(\nb_H^{++})^2 +
A_{(+)}\f{\alpha^2_H}{\sB_H - \alpha_H^2 \Omega^2} A_{(-)}\Big] \nn
\\
 &&- \f{i}2\Tr_{(4,0)}\Big(1+\f{\alpha^2_H\Omega^2}{\sB_H - \alpha^2_H
 \Omega^2}\Big) -\f{i}2 \Tr\ln (\nb_H^{++})^2\,.
 \eea

Then we consider the contribution from the quantum hypermultiplet in
\eq{1loop}. For on-shell superfields  the covariant harmonic
derivative $\nb^{++}_H$ commutes with $\sB_H$, but it is not true
for the operator $\sB_H-\alpha_H^2 \Omega^2$. Moreover,  the
operators $ A_{(\pm)}$ also contain background hypermultiplet and as
a consequence do not commute with $\sB_H -\alpha^2_H\Omega^2$ even
for the constant on-shell background superfields
 \bea
 \f{i}2 \Tr \ln\Big[(\nb_H^{++})^2 +
A_{(+)}\f{\alpha^2_H}{\sB_H - \alpha_H^2 \Omega^2} A_{(-)}\Big] =
 \f{i}2 \Tr \ln\Big[(\nb_H^{++})^2 +(\nb_H^{++})^2
\f{\alpha^2_H \Omega^2 }{\sB_H - \alpha_H^2 \Omega^2}
 + \ldots\Big], \label{Qomega}
 \eea
where dots stand for terms involving the harmonic derivative of
the hypermultiplet,  $D^{++}\Omega$.  Our aim is to demonstrate that the
$\cN=(1,1)$ invariant action \eq{I1} can be evaluated as the leading
contribution  to the one-loop effective action $\Gamma^{(1)}_{\rm
lead}$ \eq{1loop2}. The action \eq{I1} contains only the gauge
superfield strength $W^{+a}$ and $\Omega$ without terms
$D^{++}\Omega, \, D^-_a \Omega$ and $D^-_a W^{+b}$. Hence we will
systematically neglect such terms in our computations. In this case
the contributions from ghosts in \eq{1loop} and  the second term in
\eq{det} are canceled by the corresponding terms in \eq{Qomega}, and
so $\Gamma^{(1)}_{\rm high}$ collects terms with $D^{++}\Omega$ and
spinorial derivatives of the background superfields only. Thus in what follows the
contribution $\Gamma^{(1)}_{\rm high}$ will be ignored.

Computation of the expression \eq{1loop2} repeats the analogous one
in the four-dimensional case \cite{KM}. Both terms in \eq{1loop2}
contain harmonic singularities in the coincident points limit.
According to the analysis of \cite{KM}, the well-defined expression for the
contribution $ \Gamma^{(1)}_{\rm lead}$ to the one-loop effective
action reads\footnote{We have to note that the harmonic derivative
commutes with the covariant d'Alembertian on shell. But it is not
the case for the operator $\sB_H- \alpha_H^2\Omega^2$. Indeed,
$[\sB_H- \alpha_H^2\Omega^2,\nb_H^{++}]\sim D^{++}\Omega$. However,
as was mentioned above, we omit all such terms since they provide
next-to-order corrections to the leading low-energy approximation.}
 \bea
 \Gamma^{(1)}_{\rm lead}=-\frac{i}{2} \Tr \int_{0}^{\infty}
\frac{ d(is)}{(is)}e^{is(\sB_H- \alpha_H^2\Omega^2)}
\Pi^{(2,2)}_{\rm T} \,, \label{G2_2}
 \eea
where $\Pi^{(2,2)}_{\rm T}(\zeta_1,u_1; \zeta_2,u_2)$ is the projector
on the space of covariantly analytic transverse superfields
 \be
 \Pi^{(2,2)}_{\rm T} (1|2) = \d_A^{(2,2)}(1|2) -
\nb^{++}_{1} \nb^{++}_2 G^{(0,0)} (1|2) \label{Pi}\,.
 \ee
The Green function $G^{(0,0)}(\zeta_1,u_1; \zeta_2,u_2)$ satisfies
the equation
  \be
(\nb_1^{++})^2 G^{(0,0)}(1|2) = - \d_A^{(4,0)}(1|2)\,,
 \ee
and it  can be given explicitly \cite{GIOS} as
 \be
 G^{(0,0)}(\zeta_1,u_1; \zeta_2,u_2) =
 \f{(D_1^+)^4 \,(D_2^+)^4}{\sB_1}\delta^{14}(z_1-z_2) \f{(u^-_1 u^-_2)}{(u^+_1 u^+_2)^3}
 \label{G0}\,.
 \ee
By explicit calculation one can show that in our case of the on-shell
background the projector \eq{Pi} acquires the simple form
 \bea
 \Pi^{(2,2)}_{\rm T} = -\f{(D^+_1)^4}{\sB_1}
 \Big\{(\nb^-_1)^4(u^+_1u^+_2)^2
-\Omega_1^{--}(u^-_1u^+_2)(u^+_1u^+_2) + \sB_{1}(u^-_1u^+_2)^2\Big\}
\d^{14}(z_1-z_2)\,. \label{Pi2}
 \eea
We substitute the expression \eq{Pi2} for $\Pi^{(2,2)}_{\rm T}$
in the one-loop contribution $ \Gamma^{(1)}_{\rm lead}$ \eq{G2_2}
and take the coincident-harmonic points limit $u_2 \to u_1$. We see
that only the third term in \eq{Pi2} survives in this limit. Thus we
have
 \bea
\Gamma^{(1)}_{\rm lead} = -\f{i}{2}\tr\int d \zeta^{(-4)}_1
\int_0^\infty \f{d(is)}{(is)}  e^{is (\sB_H- \alpha_H^2\Omega^2)
}(D^{+}_1)^4 \d^{14} (z_1-z_2)\big|_{2=1}\,. \label{1loop3}
 \eea
The trace over matrix indices in \eq{1loop3} is reduced to a
sum over non-zero roots $\alpha_H$, taking $H =
\tfrac{1}{\sqrt{N(N-1)}}{\rm diag}(1,..,1, 1-N)$. In order to get rid of
the Grassmann delta function by using the identity $ (D^+)^4 (D^-)^4
\delta^8(\theta_1-\theta_2)\big|_{2=1}~=~1$, we collect the fourth
power of  derivative $D^-_a$ from the exponent in \eq{1loop3}. Then
we pass to the momentum representation and calculate the integral over
proper-time $s$. Finally we obtain
 \bea
\Gamma^{(1)}_{\rm lead} = \f{N-1}{(4\pi)^3}\int d \zeta^{(-4)}\,
\f{(W^+)^4}{\Omega^2}\,. \label{1loop4}
 \eea
As expected, the leading low-energy contribution \eq{1loop4}  to
the effective action in the model \eq{S0} is just the $\cN=(1,1)$
invariant $I_1$ \eq{I1}. The coefficient $c$ now takes the precise value
 \bea
 c=\f{N-1}{(4\pi)^3}\,. \label{c}
 \eea
The expression fore $c$ is similar to that in the four-dimensional $\cN=4$ SYM theory
(see, e.g., \cite{KU} and references therein). In the bosonic sector
the effective action \eq{1loop4} has the structure
 \bea
\Gamma^{(1)}_{\rm bos} \sim \int d^6 x\, \f{F^4}{\phi^2}\bigg(1+
\f{\phi^{(ij)}\phi_{(ij)}}{\phi^2}+\ldots\bigg)\,,
 \eea
where $F^4 = F_{MN} F^{MN} F_{PQ} F^{PQ} - 4 F^{NM} F_{MR} F^{RS}
F_{SN}$ and $F_{MN}$ is the abelian gauge field strength.


\section{Conclusions}

In this paper we have studied the quantum aspects of the
six-dimensional $\cN=(1,1)$ super-Yang-Mills theory. We formulated
the model in $6D, \,\cN=(1,0)$ harmonic superspace in terms of
$\cN=(1,0)$ gauge multiplet and $\omega$-hypermultiplet, all being
in the adjoint representation of gauge group $SU(N)$. By
construction, the theory possesses the manifest $\cN=(1,0)$
supersymmetry and an additional non-manifest $\cN=(0,1)$ one.

We studied the effective action of $6D, \,\cN=(1,0)$ SYM theory in
the framework of the  background field method. There was considered
the special case of the slowly varying background superfields which
break the initial gauge symmetry $SU(N)$ down to $SU(N-1)\times
U(1)\subset SU(N)$ and are subject to the free classical equations
of motion. We provided a general analysis of possible $\cN=(1,1)$
invariants which can be constructed out of the background gauge
superfield strength and hypermultiplet. We argued that one of these
invariants  can be treated as the leading low-energy contribution to
the one-loop effective action of $\cN=(1,1)$ SYM theory.

It is instructive to compare our results on the one-loop low-energy
effective action in $6D, \cN=(1,1)$ SYM theory with the recent
activity on calculating the one-loop on-shell amplitudes in the same
theory \cite{Amplitudes1,Amplitudes2,Amplitudes3}. The four-point
amplitude agrees with the $F^4$ component term in the effective
action. However, unlike the amplitudes,  the result for the
effective action in our paper was derived  in a closed superfield
form. The four-point amplitude in
\cite{Amplitudes1,Amplitudes2,Amplitudes3} does not allow to
directly restore this effective action. The point is that the
effective action (\ref{1loop4}) contains the hypermultiplet whose
scalar serves as a natural infrared regulator. The amplitudes were
calculated in the gauge multiplet sector only, that is not
sufficient for deriving the true full effective potential.

We have to note that even in the one-loop approximation there might
exist more complicated contributions to the effective action, which
can be expected on the basis of a general analysis. One can
consider, e.g., a $6D$ analog of supersymmetric Heisenberg-Euler
type effective action in $\cN=(1,1)$ SYM theory. Also, it should be
noticed that we studied only those contributions to the effective
action which contain no harmonic derivatives of the hypermultiplet superfield. The
contributions involving such derivatives were beyond the scope of
our consideration. It would be interesting to analyze such
contributions and  to consider a more general class of the background
superfields. Namely, it is tempting to find a way to consider the
complete effective action with the whole dependence on the harmonic
derivatives of the background hypermultiplet included. In this way
we expect to obtain the complete $\cN=(1,1)$ supersymmetric
quantum effective action possessing both explicit $\cN=(1,0)$ and
hidden $\cN=(0,1)$ supersymmetries. We hope to address these issues
in the forthcoming works.


\section*{Acknowledgments}

\noindent We thank D.I. Kazakov for useful discussion. This research
was supported in part by Russian Ministry of Education and Science,
project No. 3.1386.2017. I.L.B and E.A.I  are thankful to the RFBR
grant, project No 18-02-01046. I.L.B and B.S.M are grateful to RFBR
grant, project No. 18-02-00153. The work of B.S.M. was carried out
in part within the framework of Tomsk Polytechnic University
Competitiveness Enhancement Program.




\begin{thebibliography}{99}

\bibitem{BUCH1} E.I. Buchbinder, B.A. Ovrut, I.L. Buchbinder,
E.A. Ivanov, S.M. Kuzenko, {\it Low-energy effective action in N=2
supersymmetric field theories}, Phys. Part. Nucl. {\bf 32} (2001)
641-674.
%
\bibitem{BUCH2}
I.L. Buchbinder, E.A. Ivanov, N.G. Pletnev, {\it Superfield approach
to the construction of effective action in quantum field theory with
extended supersymmetry}, Phys. Part. Nucl. {\bf 47} (2016) 291-369.
%
\bibitem{BUCH3}
I.L. Buchbinder, E.A. Ivanov, I.B. Samsonov, {\it The low-energy N=4
SYM effective action in diverse harmonic superspaces}, Phys. Part.
Nucl. {\bf 48} (2017) 333-388, {\tt  arXiv:1603.02768 [hep-th]}.
%
\bibitem{GK}
A. Giveon, D. Kutasov, {\it Brane dynamics and gauge theory}, Rev.\
Mod.\ Phys.\ {\bf 71} (1999) 981-1084, {\tt arXiv:hep-th/9802067}.
%
\bibitem{BKLS}
R. Blumenhagen, B. K\"orst, D. L\"ust, S. Stieberger, {\it
Four-dimensional String Compactifications with D-Branes,
Orientifolds and Fluxes}, Phys. Repts {\bf 445} (2007) 1-193 {\tt
arXiv:hep-th/0610327}.
%
\bibitem{GKPR}
F. Gonzalez-Rey, B. Kulik, I.Y. Park and M. Ro\u{c}ek, {\it Self-dual
effective action of $\cN = 4$ super-Yang-Mills}, Nucl. Phys. B {\bf
544} (1999) 218-242, {\tt arXiv:hep-th/9810152}.
%
\bibitem{PvU}
V.~Periwal, R.~von Unge, {\it Accelerating D-branes}, Phys.\ Lett.\
B {\bf 430} (1998) 71-76, { \tt arXiv:hep-th/9801121}.
%
\bibitem{G-rR}
F. Gonzalez-Rey, M. Ro\u{c}ek, {\it Nonholomorphic $\cN = 2$ terms
in $\cN = 4$ SYM: 1-loop calculation in $\cN = 2$ superspace},
Phys.\ Lett.\ B {\bf 434} (1998) 303-311, {\tt
arXiv:hep-th/9804010}.
%
\bibitem{BBKO}
I.L. Buchbinder, E.I. Buchbinder, S.M. Kuzenko, B.A. Ovrut, {\it The
Background Field Method for N=2 Super Yang-Mills Theories in
Harmonic Superspace}, Phys.\ Lett.\ B {\bf 417} (1998) 61-71, {\tt
arXiv:hep-th/9704214}.
%
\bibitem{BK}
I.L. Buchbinder, S.M. Kuzenko, { \it Comments on the background
field method in harmonic superspace: Non-holomorphic corrections in
$\cN = 4$ SYM}, Mod.\ Phys.\ Lett.\ A {\bf 13} (1998) 1623-1636,
{\tt arXiv:hep-th/9804168}.
%
\bibitem{BBK}
E.I. Buchbinder, I.L. Buchbinder, S.M. Kuzenko, {\it Non-holomorphic
effective potential in $\cN = 4\,$ $ SU(n)$ SYM}, Phys.\ Lett.\ B
{\bf 446} (1999) 216-223, {\tt arXiv:hep-th/9810239}.
%
\bibitem{LvU}
D.A. Lowe, R. von Unge, {\it Constraints on higher derivative
operators in maximally supersymmetric  gauge theory}, JHEP {\bf
9811} (1998) 014, {\tt arXiv:hep-th/9811017}.
%
\bibitem{BKT}
I.L. Buchbinder, S.M. Kuzenko, A.A. Tseytlin, {\it On Low-Energy
Effective Actions in {\cN=2,4} Superconformal Theories in Four
Dimensions}, Phys.\ Rev.\ {\bf D62} (2000) 045001, {\tt
arXiv:hep-th/9911221}.
%
\bibitem{KM} S.M. Kuzenko, I.N. McArthur, {\it Effective action of
{\cN=4} super Yang–Mills: N = 2 superspace approach}, Phys.\ Lett.\
{\bf B506} (2001), 140-146, {\tt arXiv:hep-th/0101127}.
%
\bibitem{Kuz01} S.M. Kuzenko, I.N. McArthur, {\it Hypermultiplet effective action:
$\cN=2$ superspace approach}, Phys.\ Lett.\ B {bf 513} (2001)
213-222, {\tt arXiv:hep-th/0105121}.
%
\bibitem{BI}  I.L. Buchbinder, E.A. Ivanov, {\it Complete $\cN =
4$ structure of low-energy effective action in $\cN = 4$
super-Yang-Mills theories}, Phys. Lett. B {\bf 524} (2002) 208-216,
{\tt arXiv:hep-th/0111062}.
%
\bibitem{BIP} I.L. Buchbinder, E.A. Ivanov, A.Yu. Petrov,
{\it Complete low-energy effective action in $\cN = 4$ SYM: A direct
$\cN = 2$ supergraph calculation}, Nucl. Phys. B {\bf 653} (2003)
64-84, {\tt arXiv:hep-th/0210241}.
%
\bibitem{BPT}
I.L. Buchbinder, A.Yu. Petrov, A.A. Tseytlin, {\it Two-loop $\cN =
4$ super Yang Mills effective action and interaction  between
D3-branes}, Nucl.\ Phys.\ B {\bf 621} (2002) 179-207, {\tt
arXiv:hep-th/0110173}.
%
\bibitem{KU} S.M.Kuzenko, {\it Self-dual effective action of $\cN = 4$ SYM revisited
}, JHEP {\bf 0503} (2005) 008, {\tt arXiv:hep-th/0410128}.
%
\bibitem{BP07}
I.L. Buchbinder, N.G. Pletnev, {\it Hypermultiplet dependence of
one-loop effective action in the $\cN = 2$ superconformal theories},
JHEP {\bf 0704} (2007) 096, {\tt arXiv:hep-th/0611145}.
%
\bibitem{T}
A.A. Tseytlin, {\it On non-abelian generalization of Born-Infeld
action in string theory}, Nucl.\ Phys.\ B {\bf 501} (1997) 41-52,
{\tt arXiv:hep-th/9701125}.
%
\bibitem{Sevrin1} L. De Fosse, P. Koerber, A. Sevrin, {\it The
uniqueness of the Abelian Born-Infeld Action}, Nucl.\ Phys.\ B {\bf
603} (2001) 413-426, {\tt arXiv:hep-th/010305}.
%
\bibitem{Sevrin2}E. Bergshoeff, A. Bilal, M. de Roo, A. Sevrin, {\it Supersymmetric
non-abelian Born-Infeld revisited}, JHEP {\bf 0107} (2001) 029, {\tt
arXiv:hep-th/0105274}.
%
\bibitem{Sevrin3}P. Korber, A. Sevrin, {\it The non-abelian open superstring
effective action through order $\alpha^{'3}$}, JHEP {\bf 0110}
(2001) 003, {\tt arXiv:hep-th/0108169}.
%
\bibitem{Sevrin4}P. Korber, A. Sevrin, {\it The non-abelian D-brane
effective action through order $\alpha^{'4}$}, JHEP {\bf 0210}
(2002) 046, {\tt arXiv:hep-th/0208044}.
%
\bibitem{DHHK}
J.M. Drummond, P.J. Heslop, P.S. Howe, S.F. Kerstan, {\it Integral
invariants in $\cN=~4$ SYM and the effective action for coincident
D-branes}, JHEP {\bf 0308} (2003) 016, {\tt arXiv:hep-th/0305202}.
%
\bibitem{GTKK}
T.W. Grimm, Tae-Won Ha, A. Klemm, D. Klevers, {\it The D5-brane
effective action and superpotential in $\cN=1$ compactifications},
Nucl.\ Phys.\ B {\bf 816} (2009) 139-184, {\tt arXiv:0811.2996
[hep-th]}.
%
\bibitem{FT} E.S. Fradkin, A.A. Tseytlin, {\it Quantum properties of higher dimensional and
dimensionally reduced supersymmetric theories}, Nucl. Phys. B {\bf
227} (1983) 252-290.
%
\bibitem{MarSag1}
N. Markus, A. Sagnotti, {\it A test of finiteness predictions for
supersymmetric theories}, Phys. Lett. {\bf B 135} (1984) 85-90.
%
\bibitem{MarSag2}
N. Markus, A. Sagnotti, {\it The Ultraviolet Behavior of ${\cal
N}=4$ Yang-Mills and Power Counting of Extended Superspace}, Nucl.
Phys., {\bf B 256} (1985) 77-108.
%
\bibitem{HS}
P.S. Howe, K.S. Stelle, {\it Ultraviolet Divergences in Higher
Dimensional Supersymmetric Yang-Mills Theories}, Phys. Lett. B {\bf
137} (1984) 175-180.
%
\bibitem{HS1}
P.S. Howe, K.S. Stelle, {\it Supersymmetry counterterms revisited},
Phys. Lett. B {\bf 554} (2003) 190-196, {\tt arXiv:hep-th/0211279}.
%
\bibitem{BHS}
G. Bossard, P.S. Howe, K.S. Stelle, {\it The Ultra-violet question
in maximally supersymmetric theories}, Gen. Relat. Grav. {\bf 41}
(2009) 919, {\tt arXiv:0901.4661 [hep-th]}.
%
\bibitem{BHS1}
G. Bossard, P.S. Howe, K.S. Stelle, {\it A Note on the UV behaviour
of maximally supersymmetric Yang-Mills theories}, Phys. Lett. B {\bf
682} (2009) 137-142, {\tt arXiv:0908.3883 [hep-th]}.
%
\bibitem{Bork}
L.V. Bork, D.I. Kazakov, M.V. Kompaniets, D.M. Tolkachev, D.E.
Vlasenko, {\it Divergences in maximal supersymmetric Yang-Mills
theories in diverse dimensions}, JHEP {\bf 1511} (2015) 059, {\tt
arXiv:1508.05570 [hep-th]}.
%
\bibitem{BIMS-a}
I.L. Buchbinder, E.A. Ivanov, M.B. Merzlikin, K.V. Stepanyantz, {\it
One-loop divergences in the $6D, {\cal N}=(1,0)$ Abelian gauge
theory}, Phys. Lett. B {\bf 763} (2016) 375-381, {\tt
arXiv:1609.00975 [hep-th]}.
%
\bibitem{BIMS-b}
I.L. Buchbinder, E.A. Ivanov, B.S. Merzlikin, K.V. Stepanyantz, {\it
One-loop divergences in 6D, N=(1,0) SYM theory}, JHEP {\bf 1701}
(2017) 128, {\tt arXiv:1612.03190 [hep-th]}.
%
\bibitem{BIMS-c}
I.L. Buchbinder, E.A. Ivanov, B.S. Merzlikin, K.V. Stepanyantz, {\it
Supergraph analysis of the one-loop divergences in $6D$, ${\cal N} =
(1,0)$ and ${\cal N} = (1,1)$ gauge theories}, Nucl.\ Phys.\ B {\bf
921} (2017) 127 - 158, {\tt arXiv:1704.02530 [hep-th]}.
%
\bibitem{BIMS-d}
I.L. Buchbinder, E.A. Ivanov, B.S. Merzlikin, K.V. Stepanyantz, {\it
On the two-loop divergences of the 2-point hypermultiplet
supergraphs for $6D, \cN=(1,1)$ SYM theory}, Phys.\  Lett.\ B {\bf
778} (2018) 252 - 255, {\tt arXiv:1711.11514 [hep-th]}.
%
\bibitem{GIKOS}
A. Galperin, E. Ivanov, S. Kalitzin, V. Ogievetsky, E. Sokatchev,
{\it Unconstrained $N=2$ matter, Yang-Mills and supergravity
theories in harmonic superspace}, Class. Quantum Grav. {\bf 1}
(1984) 469-498.
%
\bibitem{GIOS}
A. S. Galperin, E. A. Ivanov, V. I. Ogievetsky, E. S. Sokatchev,
{\it Harmonic Superspace}, Cambridge University Press, Cambridge,
2001, 306 p.
%
\bibitem{BIS}G. Bossard, E. Ivanov, A. Smilga, { \it Ultraviolet behaviour
of 6D supersymmetric Yang-Mills theories and harmonic superspace},
JHEP {\bf 1512} (2015) 085, {\tt arXiv:1509.08027 [hep-th]}.
%
\bibitem{HSW}
P.S. Howe, K.S. Stelle, P.C. West, {\it N = 1 d = 6 harmonic
superspace}, Class. Quant. Grav. {\bf 2} (1985) 815.
%
\bibitem{Z}
B.M. Zupnik, {\it Six-dimensional supergauge theories in the
harmonic superspace}, Sov. J. Nucl. Phys. {\bf 44} (1986) 512.
%
\bibitem{C1}
M. Cederwall, {\it Pure spinor superfields-an overview} , Springer
Proc. Phys. {\bf 153} (2013) 61, {\tt arXiv:1307.1762 [hep-th]}.
%
\bibitem{C2}
M. Cederwall, {\it Pure spinor superspace action for $D=6, \cN=1$
super-Yang-Mills theory}, {\tt arXiv:1712.02284 [hep-th]}.
%
\bibitem{book}
I.L. Buchbinder and S.M. Kuzenko, {\it Ideas and methods of
supersymmetry and supergravity: Or a walk through superspace},
Bristol, UK: IOP (1998) 656 p.
%
\bibitem{S_98} N. Seiberg,
{\it Notes on theories with 16 supercharges}, Nucl.\ Phys.\ Proc.\
Suppl.\ {\bf 67} (1998) 158-171, {\tt arXiv:hep-th/9705117}.
%
\bibitem{Amplitudes1}
T. Dennen, Yu-tin Huang, W, Siegel, {\it Superwistor space for 6D
maxiam super Yang-Mills}, JHEP {\bf  1004} (2010) 127, {\tt
arXiv:0910.2688 [hep-th]}.
%
\bibitem{Amplitudes2}
A. Brandhuber, D. Korres, D. Koschade, G. Travaglini, {\it One-loop
Amplitudes in Six-Dimensional (1,1) Theories from Generalised
Unitarity}, JHEP {\bf 1102} 2011 077, {\tt arXiv:1010.1515
[hep-th]}.
%
\bibitem{Amplitudes3}
L.V. Bork, D.I. Kazakov, D.E. Vlasenko, {\it On the amplitudes in
$\cN=(1,1) D=6$ SYM}, JHEP {\bf 11}(2013)065, {\tt arXiv:1308.0117
[hep-th]}.


\end{thebibliography}
\end{document}